\documentclass{article}

     \usepackage[final]{neurips_2023}

\usepackage[utf8]{inputenc} 
\usepackage[T1]{fontenc}    
\usepackage{hyperref}       
\usepackage{url}            
\usepackage{booktabs}       
\usepackage{amsfonts}       
\usepackage{nicefrac}       
\usepackage{microtype}      
\usepackage{xcolor}         
\usepackage{graphicx}      

\title{Skilful Precipitation Nowcasting Using NowcastNet}

\author{%
  Ajitabh Kumar\\
  Enrflo\\
  Noida, UP 201301 India \\
  \texttt{ajitabh.kumar@enrflo.com} \\
}

\begin{document}

\maketitle

\begin{abstract}
  Designing early warning system for precipitation requires accurate short-term forecasting system. Climate change has led to an increase in frequency of extreme weather events, and hence such systems can prevent disasters and loss of life. Managing such events remain a challenge for both public and private institutions. Precipitation nowcasting can help relevant institutions to better prepare for such events as they impact agriculture, transport, public health and safety, etc. Physics-based numerical weather prediction (NWP) is unable to perform well for nowcasting because of large computational turn-around time. Deep-learning based models on the other hand are able to give predictions within seconds. We use recently proposed NowcastNet, a physics-conditioned deep generative network, to forecast precipitation for different regions of Europe using satellite images. Both spatial and temporal transfer learning is done by forecasting for the unseen regions and year. Model makes realistic predictions and is able to outperform baseline for such a prediction task. Our solution was ranked third in both nowcasting and transfer learning leaderboards.
\end{abstract}

\section{Introduction}

Precipitation forecasting is important for planning day-to-day activities as well as mitigating disasters. Climate change brings new set of challenges for government as extreme weather events become more frequent. Nowcasting for 0-8 hour range can help planners navigate these challenges but traditional NWP based approaches have relatively large turn around time. State-of-the-art NWP methods typically advect precipitation fields with radar-based wind estimates, and fail to capture convective initiations [1, 2]. Recently deep learning based approaches have been used in nowcasting as they can yield predictions within seconds using available satellite and radar images. Different studies have shown that deep learning based approaches can even outperform NWP in nowcasting problem.

Weather4Cast $2023$ challenge is organized in conjunction with NeurIPS $2023$ conference for developing and comparing different deep learning approaches for solving such a nowcasting problem. Training data is available for seven different regions of Europe for two years (i.e. $2019$ and $2020$). Goal of challenge is to predict the precipitation for next $8$ hours given satellite images for last $1$ hour [3]. Next, spatial transfer learning problem involves forecasting precipitation for another three regions of Europe for which no training data is provided. Finally, temporal transfer learning problem involves predicting precipitation for all ten regions for the third year (i.e. $2021$).

Precipitation forecasting is challenging as it is highly heterogeneous event and requires dense prediction. Smooth and blurry prediction can easily diminish forecast usability, and so rain intensities must be captured. Core challenge of this competition is to create rain movie at $15$ minute interval for the next $8$ hour. NowcastNet incorporates both physics and data to make realistic precipitation forecast [4, 5].

\section{Method}

In this section, we first discuss available training data and then we discuss the deep learning architecture used.

\subsection{Dataset}

Ground-radar reflectivity measurements are used to calculate pan-European composite rainfall rates by the Operational Program for Exchange of Weather Radar Information (OPERA) radar network. Even though these are more precise, accurate, and of higher resolution than satellite data, they are expensive to obtain and not available in many parts of the world. Hence, goal of this competition is to develop deep learning based models to predict rain rates from radiations measured by geostationary satellites, which are operated by the European Organisation for the Exploitation of Meteorological Satellites (EUMETSAT).

\begin{figure}
  \centering
  \includegraphics[width=0.8\linewidth]{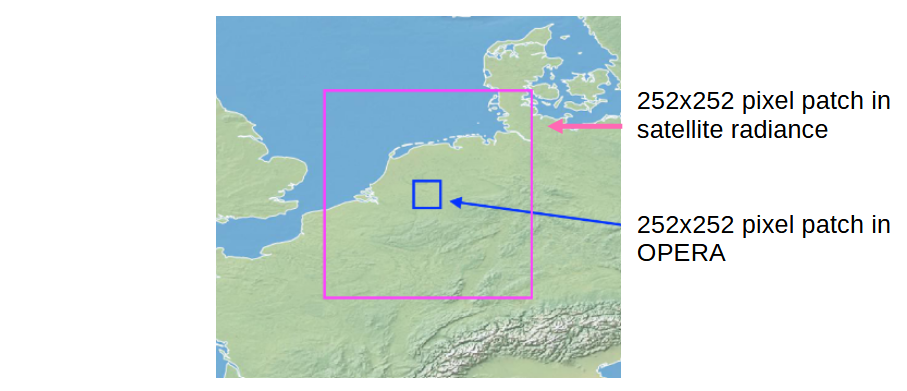}
  \caption{Central area of interest within larger context area for a region under study}
  \label{context}
\end{figure}

Satellite based images with $11$ different channels are available for larger context region of area $3024\times3024$ km. Each image contains data point for every grid size of $12\times12$ km, and hence these images are of $252\times252$ dimensions. Channels denote slightly noisy satellite radiance covering so-called $2$ visible (VIS), $2$ water vapor (WV), and $7$ infrared (IR) bands. Precipitation data on the other hand is available for the central $504\times504$ km area with a grid size of $2\times2$ km. Thus, even precipitation images are of $252\times252$ dimension. Figure~\ref{context} shows the larger context area with satellite data and the central area with OPERA based precipitation data for one of the regions. Input data is cleaned in the pre-processing step in which all unavailable values are replaced with $0$. Input values are normalized so that different channels could be used together.

\subsection{NowcastNet}

\begin{figure}
  \centering
  \includegraphics[width=0.8\linewidth]{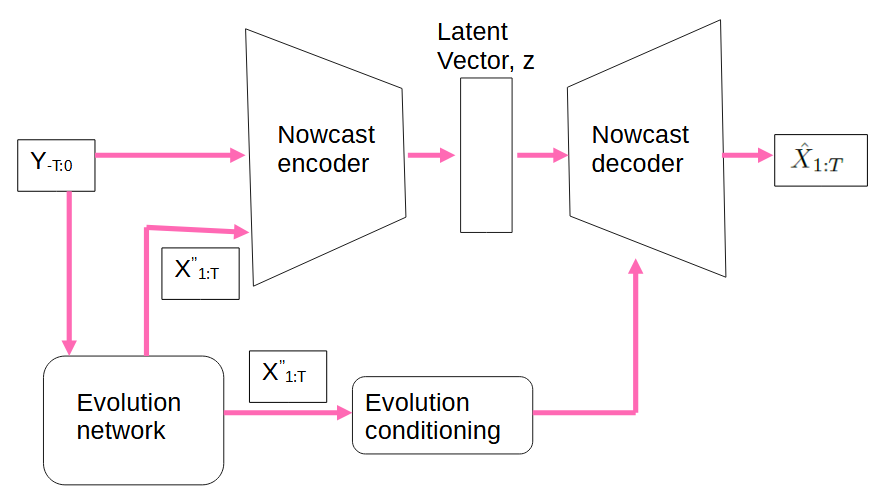}
  \caption{NowcastNet architecture: a physics conditional deep generative network}
  \label{nowcastnet}
\end{figure}

Deep learning core task is to predict high-resolution ($2\times2$ km) precipitation values using low resolution ($12\times12$ km) satellite images. NowcastNet has two main modules: an evolution network which learns underlying physics using neural operators, and a generative network which makes final prediction using context from the evolution network (Figure~\ref{nowcastnet}). Noise projector is added in generative network which transforms the latent vector into required spatial size. Thus, the model includes a deterministic evolution network parameterized by $\phi$ and a stochastic generative network parameterized by $\theta$. Nowcasting procedure is thus physics-conditional generation from latent random vectors z, given as:

\begin{equation}
P(\hat{X}_{1:T}|Y_{-T_{0}:0}, \phi;\theta) = \int P(\hat{X}_{1:T}|Y_{-T_{0}:0}, \phi(Y_{-T_{0}:0}),z;\theta)P(z)dz
\end{equation}

where $Y_{-T_{0}:0}$ denotes given past satellite images and $\hat{X}_{1:T}$ denotes predicted precipitation image.

\begin{figure}
  \centering
  \includegraphics[width=0.8\linewidth]{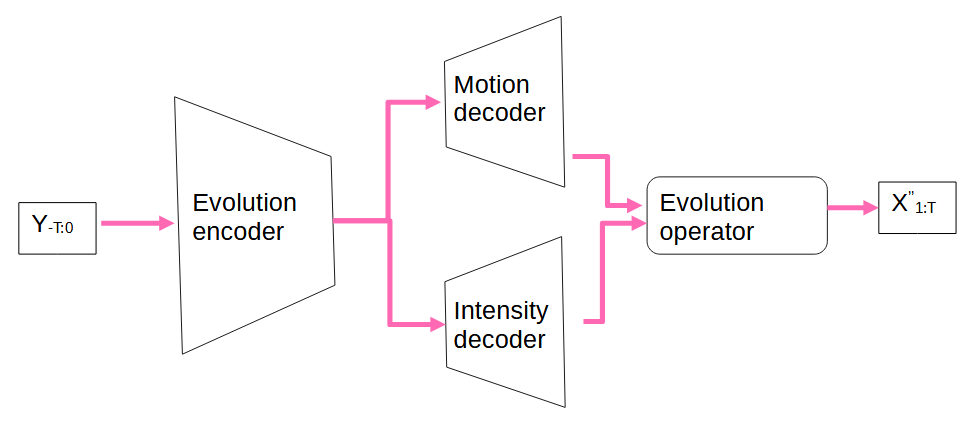}
  \caption{Evolution network and operator}
  \label{evolution}
\end{figure}

Evolution network is a two path U-Net with shared evolution encoder for learning context representation (Figure~\ref{evolution}). Motion decoder learns the two-dimensional velocities for x- and y-directions, while intensity decoder learns intensity residuals for the desired times. Spectral normalization technique is applied for each convolutional layer. Input and output values are concatenated on temporal dimension for skip connection.

Generative network conditions on the evolution network predictions $X^{''}_{1:T}$, and takes low-resolution past satellite images as input $X_{-T_{0}:0}$. It generates final precipitation field from latent random vectors $z$. The backbone of this network is also a U-Net encoder-decoder structure. The encoder has same structure as the evolution encoder, and takes as input concatenation of $X_{-T_{0}:0}$ and $X^{''}_{1:T}$. The decoder takes contextual input from encoder, along with the transformation of the latent Gaussian vector $z$. Noise projector takes this latent vector $z$ as input and transforms it to the same spatial size as the encoder context.

\section{Results}

Original input data consists of low-resolution satellite image with $11$ different channels for a larger context area, while high-resolution prediction is done for a smaller central area. We use Adam optimizer and cosine annealing learning rate with linear warm-up. Mean squared error is used as loss function in the initial epochs, which is changed to weighted mean absolute error in the later epochs. Pixel-wise weight is calculated as $min(1+y, 24)$ where $y$ is the target precipitation rate. This helps in learning less frequent but high value precipitation events, and ensuring that network does not only fit to dominant medium to low value precipitation events.

For the core challenge, input data consists of $4$ past values at $15$-minute interval (i.e. past $1$ hour), while prediction is made for next $32$ future times at $15$-minute interval (i.e. next $8$ hours). Next for the nowcasting and transfer-learning challenge, input data consists of $4$ past values at $15$-minute interval (i.e. past $1$ hour), while prediction is made only for the next $16$ future times at $15$-minute interval (i.e. next $4$ hours). First, we use the previously learnt model and fine-tune it for the $16$ output values. Next we also train a model from scratch with only $16$ output values.

\begin{figure}
  \centering
  \includegraphics[width=0.8\linewidth]{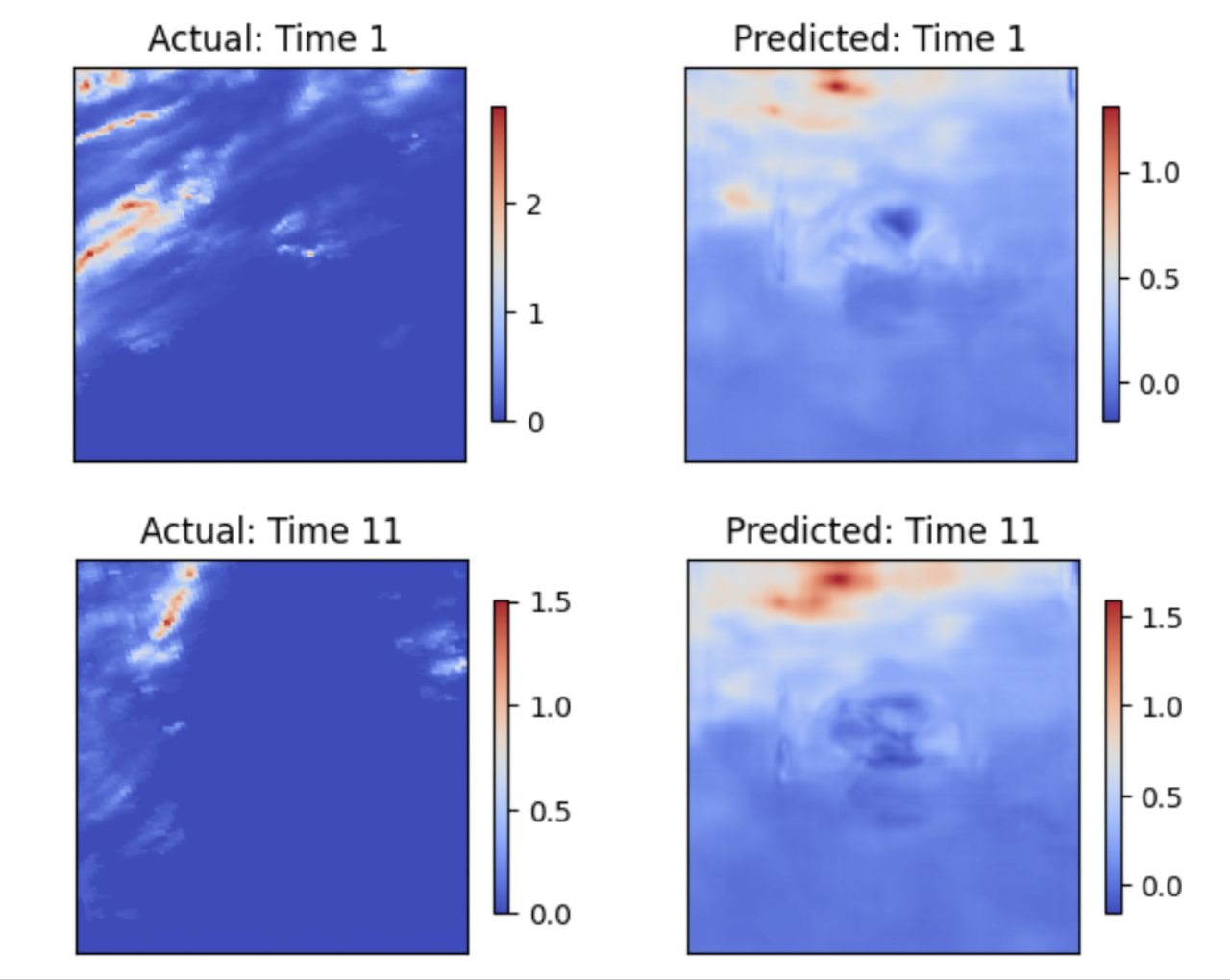}
  \caption{Actual and predicted rainfall values for different prediction times (case 1)}
  \label{case1}
\end{figure}

\begin{figure}
  \centering
  \includegraphics[width=0.8\linewidth]{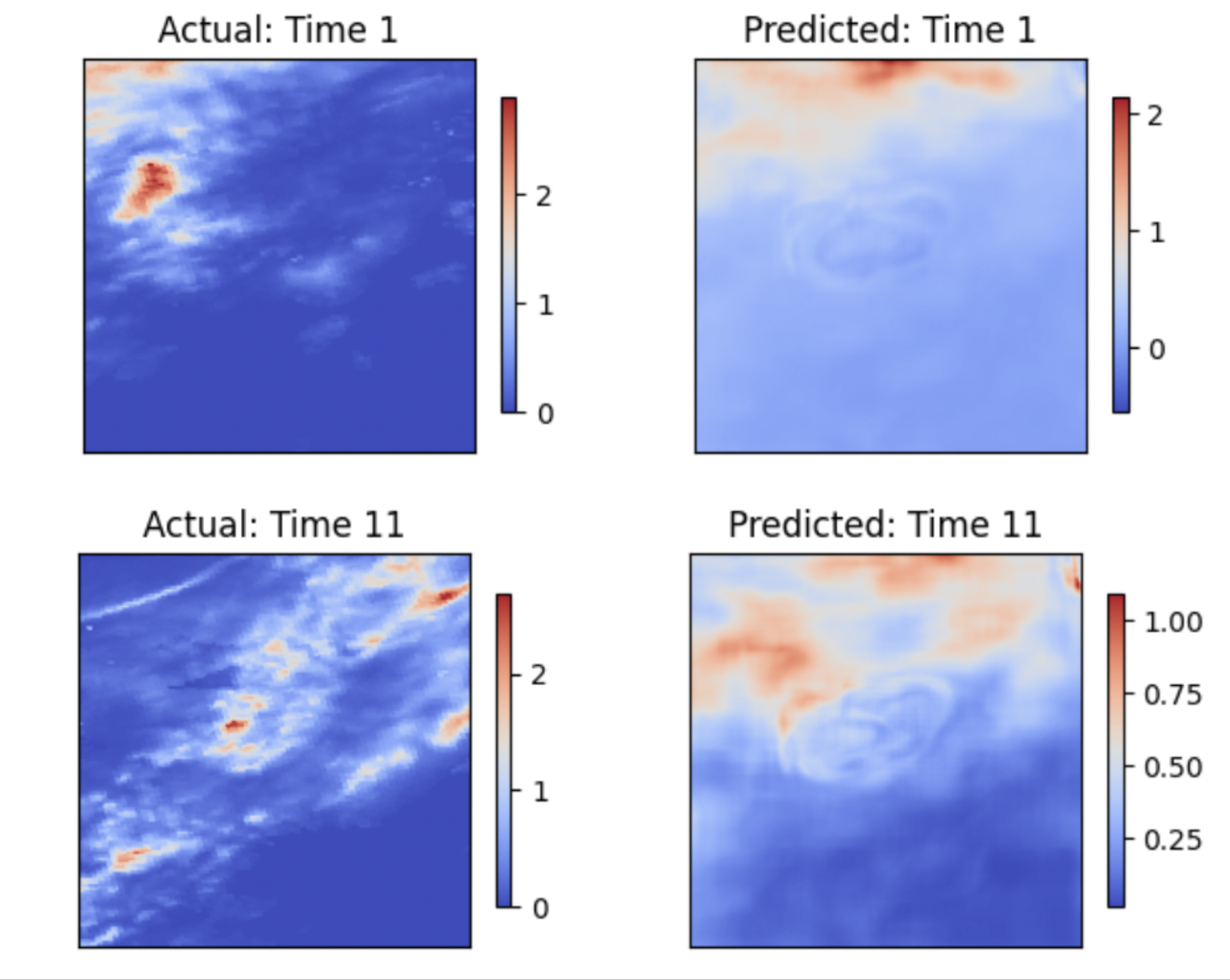}
  \caption{Actual and predicted rainfall values for different prediction times (case 2)}
  \label{case2}
\end{figure}

All different approaches yield similar result and some sample prediction images are shown next. Figures~\ref{case1} and ~\ref{case2} show the actual and predicted rainfall values for two cases. As we can see, nowcastnet based model is able to capture rain events. More improvement could be made in result by pre-processing data appropriately to capture logarithmic trend and non-stationarity in data. Nowcastnet performs well and captures both advective and convective effects using evolution and generative networks.

\section{Conclusions}

NowcastNet architecture based neural network model is trained for precipitation nowcasting. Evolution network of the model learns underlying physics using neural operators, while generative network which makes final prediction using context from the evolution network. Model is able to predict rainfall events using satellite based images.

\begin{ack}
This research is made possible by generous cloud credits provided by OVH Cloud.
\end{ack}

\section*{References}

\medskip

{
\small

[1] Asperti, A., Merizzi, F., Paparella, A., Pedrazzi, G., Angelinelli, M.\ \& Colamonaco, S.\ (2023) Precipitation nowcasting with generative diffusion models. {\it arXiv preprint arXiv:2308.06733}.

[2] Pulkkinen, S., Nerini, D., P\'erez Hortal, A.A., Velasco-Forero, C., Seed, A., Germann, U.\ \& Foresti, L.\ (2019) Pysteps: an open-source Python library  for probabilistic precipitation nowcasting. {\it Geoscientific Model Development} {\bf 12}(10):4185-4219.

[3] Pihrt, J., Raevskiy, R., Šimánek, P.\ \& Choma, M.\ (2022) WeatherFusionNet: Predicting precipitation from satellite data. {\it arXiv preprint arXiv:2211.16824}.

[4] Zhang, Y., Long, M., Chen, K.Xing, L, Jin, R., Jordan, M.I.\ \& Wang, J.\ (2023) Skilful nowcasting of extreme precipitation with NowcastNet. {\it Nature} {\bf 619}:526–532.

[5] Ravuri, S., Lenc, K., Willson, M. et al.\ (2021) Skilful precipitation nowcasting using deep generative models of radar. {\it Nature} {\bf 597}:672–677.

}

\end{document}